\documentclass[aps,prl,reprint,superscriptaddress]{revtex4-2}
\usepackage{amsmath,amssymb,bm}
\usepackage{graphicx}

\usepackage{enumerate}
\usepackage{color}
\usepackage[utf8]{inputenc} 
\usepackage[english]{babel}
\usepackage[T1]{fontenc}
\usepackage{amsfonts,amssymb,amsmath,latexsym,amsthm}
\usepackage{textcomp}
\usepackage[pdftex]{hyperref}
\usepackage{geometry}
\usepackage[toc,page]{appendix}

\begin{document}

\title{ Trapping: The “Waterfall” of Vlasov-Poisson Dynamics and the Post-transient Emergence of Drift-Independent Hole Structures in Collisionless Plasmas }

\author{Hans Schamel} 
\affiliation{Physikalisches Institut, Universität Bayreuth, 95440 Bayreuth, Germany \\email: hans.schamel@googlemail.com}
\date{\today}

\begin{abstract}

A key aspect of structure formation in collisionless plasmas is "particle trapping," or more precisely, the gap this process leaves in the theoretical description. Numerical simulations reveal this to be a transient, strongly nonlinear, and rather indeterministic process involving both the trapping of particles within the wave potential and the phase synchronization of the wave itself.
While Landau physics and phase mixing govern the dynamics prior to the onset of "trapping," the full impact of nonlinearity only becomes apparent during the trapping phase—specifically in the form of equilibrium structures, which constitute the most distinctive feature of the post-trapping phase.
Consequently, this gap—or "black box"—divides the temporal evolution into a linear Vlasov-Poisson (lVP) phase preceding the transition and a subsequent Vlasov-Poisson (VP) phase. 
This transition region gives rise to a wide spectrum of trapping scenarios, upon which the Schamel equilibria are based.

In a figurative sense, this trapping process can thus be compared to a \textbf{waterfall} in fluid mechanics, where likewise no explicit dynamic connection can be established between the region above and the region below the cascade.

Apart from the complexity of the trapping process itself—which must be viewed as a multi-scale transient phenomenon defying mathematical description—another reason for the emergence of the gap lies in the singular, chaotic behavior of particle trajectories near the separatrix, as well as the breakdown of the mean-field Vlasov-Poisson approximation in this region.
In this regime, correlations—and thus the inclusion of the BBGKY hierarchy—become necessary. These induce diffusion in phase space and result in a coarse-grained plasma in which the hole structures are embedded. A method of matched asymptotic expansions is proposed to identify the trapping scenarios relevant to these holes.



Furthermore, we report for the first time on the theoretical existence of a spatially periodic Langmuir hole—a structure for which neither a solitary limiting case nor an evolution equation exists, and which disappears if overly strict regularity requirements are imposed.
We also show that self-acceleration and the transition from a slow-electron-acoustic hole  to a fast Langmuir hole are accompanied by the release of particularly deeply trapped electrons—a process that, as expected, eludes mathematical description, since the release would involve the singular separatrix region.

\end{abstract}

\maketitle

\section{Introduction}

The formation of coherent electrostatic structures in collisionless plasmas remains a central problem in kinetic theory. While the Vlasov--Poisson system provides an exact microscopic description, its nonlinear evolution from (smooth or less smooth) initial conditions typically leads to trapping that is characterized  by a complex resonant wave-particle interaction process and  involves events  such as phase-space filamentation, folding, trapping or detrapping, separatrix crossing and stochastic trajectories (characteristics) etc. 
Coarse-graining therefore appears inevitable, leaving behind a quiescent plasma in which electrostatic equilibrium structures are embedded.\\
Two approaches to their description have been proposed: the BGK method \cite{BGK57} and the Schamel method \cite{S72}.\\
While the BGK method has achieved cult status as the first technique to demonstrate the existence of such structures, it neglects the underlying physics and treats trapped particles merely as a retrospective embellishment, without allowing for control over the configuration of these particles.\\
Schamel's method, by contrast, regards trapped particles as the fundamental cause of the nonlinearity and provides the physical basis for such structures by introducing internal degrees of freedom that effectively capture the multitude of possible trapping configurations.\\

The hypothesis presented here addresses the explicit solvability—and, in particular, the uniqueness—of the Cauchy initial value problem. This issue arises because the non-deterministic, ergodic behavior of trajectories at the moment of particle trapping disrupts the Hamiltonian flow to which the Vlasov system is normally subject.
A "black box" emerges—akin to a waterfall in fluid dynamics—within which trapping scenarios unfold that elude analytical tracking and challenge the notion of uniqueness. They can only be understood retrospectively through a comparison with Schamel’s equilibrium solutions.\\

We illustrate this using a simplified system—electron Vlasov-Poisson dynamics with immobile ions—and interpret two numerical simulations in which the concept of trapping scenarios directly influences the structures that emerge after the transient trapping phase.

\section{Vlasov-Poisson System and Schamel method}

Here we shortly repeat the method that has been introduced  by Schamel half a century ago in \cite{S72}. For simplicity  we assume immobile ions, represented by a constant ion density. This VP system consists of the Vlasov equation for electrons and   Poisson's equation and is for normalized quantities in the rest frame of the wave given by
\begin{eqnarray}
[v\partial_x + \phi'(x)\partial_v] f_e(x,v)=0 \qquad \\
\phi''(x)=\int \mathrm{d}v f_e(x,v) - 1
\end{eqnarray}

The normalization is standard, electron quantities are normalized by their thermal expressions and space and time by the Debye length and the inverse plasma frequency, respectively.

A solution of the Vlasov equation is provided by the constants of motion: $\epsilon_e=\frac{v^2}{2} -\phi$ and $ \sigma_e=\frac{v}{|v|}$  in which the sign constant refers to untrapped electrons only.
For the sake of simplicity, we choose a Maxwellian background plasma and assume a positive potential pedestal $0 \le \phi(x) \le \psi$.

 The distribution-ansatz is given  by:

\begin{eqnarray}
f_e(x,v) =C
\begin{cases}
e^{(-\frac{1}{2}(\sigma_e \sqrt{2\epsilon_e}-
\tilde v_D)^2 )}, & \epsilon_e > 0 \\
e^{-\tilde v_D^2/2)}e^{(-\beta \epsilon_e + \gamma\sqrt{-\epsilon_e})}, & \epsilon_e < 0
\end{cases}
\end{eqnarray}
where $C:=\frac{1+k_{0}^2\psi}{\sqrt{2\pi}}$. It is  known as Schamel distribution.
The first part $\epsilon_{e}>0$ refers to the free electrons, the second part $\epsilon_{e}<0$ to trapped electrons.
Regarding the distribution of trapped electrons, we consider two quite natural trapped electron configurations, which we designate as the $\beta$-trapping  and $\gamma$-trapping scenarios, respectively.
Note that a negative trapping parameter $\beta$ describes a depression in the distribution of trapped electrons.
While other, more complex configurations of trapped particles are possible, they are neglected here; some of them are discussed in \cite{S23}. Of particular interest is a jump in the distribution at the separatrix, which is addressed in \cite{S15}. They broaden the wave spectrum but simultaneously complicate the search for structures; see \cite{S23}.
Our distribution considered here is continuous.

In addition, we take into account an electric current in the plasma determined by the drift velocity $v_D$ between electrons and ions, and define the expression $\tilde v_D := |v_D - v_0|$, which incorporates the phase velocity $v_0$.

To obtain the density, we have to integrate over the entire velocity space.

For small amplitudes, $\psi<<1$, we get with $Z_r(x):=\frac{1}{\sqrt \pi} P\int dt \frac{exp(-t^2)}{t-x}$, being the real part of the complex plasma dispersion function, 
\begin{eqnarray}
n_e=1 + \frac{k_{0}^2}{2}\psi + \Gamma -\frac{1}{2}Z_r'(\frac{\tilde v_D}{\sqrt 2})\phi  - \frac{5 B}{4\sqrt\psi}\phi^{3/2} + ...
\end{eqnarray},

where we have defined:

\begin{eqnarray}
B:=\frac{16(1-\beta-\tilde v_D^2)}{15 \sqrt \pi}\exp(-\tilde v_D^2/2) \sqrt \psi \\
\Gamma:=\frac{\sqrt \pi }{2}\exp(-{\frac{\tilde v_D^2}{2}}) \gamma
\end{eqnarray}
This expression coincides e.g. with (20b) of \cite{S00}.

To solve Poisson's equation, we introduce the (provisional) pseudo-potential in formal analogy to classical mechanics:
$\mathcal V_0(\phi;v_0)$ by $\phi''(x) = n_e(\phi) - 1 =: -\mathcal V_0'(\phi;v_0)$, where in the latter the derivative refers to  $\phi$. We get
$ -\mathcal V_0'(\phi;v_0) =\frac{k_{0}^2}{2}\psi +[\Gamma-\frac{1}{2}Z_r'(\frac{\tilde v_D}{\sqrt 2}) ]\phi -\frac{5}{4\sqrt\psi}[ B \phi^{3/2} ]  $. By $\phi$ integration, assuming that $\mathcal V_0(\phi;v_0)$ vanishes at $\phi=0$ we get 

$$ -\mathcal V_0(\phi;v_0) =\frac{k_{0}^2}{2}\psi \phi  + [\Gamma-\frac{1}{2}Z_r'(\frac{\tilde v_D}{\sqrt 2}) ]\frac{\phi^2}{2}\\
 -\frac{B}{2\sqrt\psi} \phi^{5/2}  $$
 
By x-integration of  Poisson's equation it thus holds the pseudo-energy: $$ \qquad \frac{\phi'(x)^2}{2}+\mathcal V_0(\phi;v_0)=0.$$
Since at the potential maximum $\phi=\psi$ the slope of $\phi(x)$ (or the first derivative $\phi'(x)$) vanishes, we arrive directly at: $\mathcal V_0(\psi;v_0)=0$,  which is a determining equation for $v_0$.

This equation is commonly referred to as the nonlinear dispersion relation (NDR) and is:
\begin{eqnarray}  \label{eq30}
k_{0}^2 + \Gamma - \frac{1}{2}Z_r'(\frac{\tilde v_D}{\sqrt 2})  = B 
\end{eqnarray}
Its solution $v_0$ provides the first part of our problem of finding a suitable $\phi(x-v_0t)$. The second part, the determination of the profile $\phi(x)$, follows directly from the canonical form of the pseudo-energy:
\begin{eqnarray}
  \qquad \frac{\phi'(x)^2}{2}+\mathcal V(\phi)=0
\end{eqnarray}

The canonical pseudo-potential $\mathcal V(\phi)$ is thereby obtained by replacing the $v_0$-dependent part in $\mathcal V_0(\phi;v_0)$ by the NDR and is given by:
\begin{eqnarray} \label{eq32}
 -\mathcal V(\phi) =\frac{k_{0}^2}{2} \phi ( \psi-\phi)   +\frac{B}{2\sqrt\psi}\phi^2(\sqrt \psi- \sqrt\phi) 
\end{eqnarray}
Equations (\ref{eq30})-(\ref{eq32}) provide the general solution of our problem with two trapping scenarios $B$ and $\Gamma$. Note that the latter only appears in the NDR but not in the pseudo-potential. The NDR and the pseudo-potential are identical to earlier expressions such as (24),(25) or (44),(45) of \cite{S00} or (51),(52) of \cite{S23}, respectively. \\

\section{Solitary electron holes of opposite polarity, single harmonic and cnoidal holes}

For given $(k_0, B)$, (9) yields a periodic wave structure described by Jacobi elliptic functions. This includes the single harmonic wave specified for $B=0$.

A solution exists as long as the condition $-2k_0^2 \le B$ is satisfied. The lower limit $-2k_0^2 = B < 0$ corresponds to an inverted solitary electron hole, i.e. a hole with negative potential polarity (also referred to as a "Solitary Potential Dip" or SPD). This typically requires $\beta > 0$ \cite{S12}. 
The standard solitary electron hole (SEH)—that is, the one with a positive potential hill—is characterized by $k_0^2=0$ and $B>0$ and typically requires $\beta < 0$. Precisely this structure is frequently observed in numerical simulations; it has given rise to the term "hole," as there is an electron deficit in the trapped region of phase space.  Analytical expressions are derived in \cite{S12,S15}. 

Please note that neither the drift velocity $v_D$ nor the specific trapping scenario $\gamma$ affects the profile $\phi(x)$, which is obtained simply by evaluating (8) and (9). Schamel's profiles are  independent of speed and drift.

\section{On- and Off- Dispersion Branches}

We now turn to the question of which phase velocities $v_0$ are generally permissible — a question that the BGK method leaves unanswered. We restrict our analysis to a current-free plasma ($v_D=0$) and neglect the $\gamma$-trapping scenario ($\gamma=0=\Gamma_e$). Their effects can be treated in a similar manner.

In this case, the NDR simplifies to

\begin{eqnarray} 
k_{0}^2  - \frac{1}{2}Z_r'(\frac{v_0}{\sqrt 2})  = B
\end{eqnarray}
where in $B$ (5) $\tilde v_D$ is replaced by $v_0$.

Fig. 1 shows the nonlinear dispersion relation (NDR) as a multi-branched dispersion relation $\omega_0(k_0)$ for several $B$,  where $\omega_0 := k_0v_0$.

\begin{figure}[h!]
	\centering	
	\includegraphics[scale=0.3]{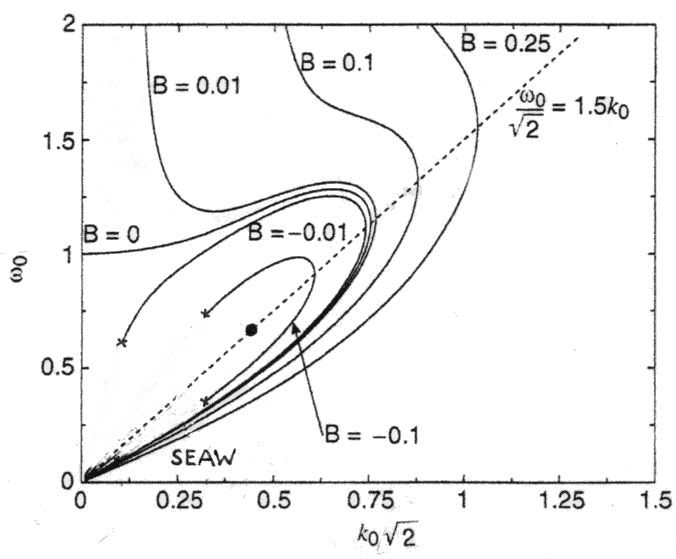}
	\caption{The NDR (11) with $ \omega_0:=k_{0}v_0$}
	\label{Fig1}
\end{figure}

Solutions with $B=0$ are referred to as on- dispersion branches, and those with $B\ne0$ as off- dispersion branches.\\

i) On-dispersion branches (thumb dispersion relation)

This curve, reminiscent of a thumb, arises when the factor $(1-\beta-v_0^2)$ in $B$ (5) equals zero; it is thus a purely nonlinear phenomenon that only appears in VP space *a posteriori*—that is, after the transient trapping process has concluded.

For long-wavelength structures ($k_0^2 << 1$), it exhibits two branches: the Langmuir branch $\omega_0 = 1 + 3k_0^2/2 + \dots$  and the slow electron-acoustic branch $\omega_0 = 1.307 k_0(1 + k_0^2 + \dots)$. (Note: The “teardrop” part of the “thumb-teardrop” dispersion relation is absent here; this feature arises from the presence of mobile ions.)

We acknowledge the existence of a single harmonic wave—note that $B=0$ in (9)—which relies on trapping and is thus unrelated to linear Vlasov dynamics. It is precisely this mode that proves to be marginally stable—in contrast to Landau's scenario—regardless of any drift \cite{S18}.

An important point is that, due to the factor $e^{-v_0^2/2}$ in (5), $B$ can effectively become zero for large values of $v_0^2$. In this case, no particle trapping occurs—or rather, none is required; we revert to the physics of the linear Vlasov-Poisson (lVP) equations and recover the undamped harmonic Langmuir wave known from standard plasma theory. This mode thus exists in both regimes and continues to play a role even after the trapping process.
Numerical simulations—both the subcritical simulations by Mandal, Sharma, and Schamel \cite{SMS17,MSS18,MSS20,SMS20} and the simulations of nonlinear Landau damping by Ouyang, Zhu, and Ng \cite{Ouyang26}—show that this mode is indeed the most energetic and manifests itself as two counter-propagating Langmuir waves.\\

ii) Off-dispersion branches

In the case $B \ne 0$, we first identify a cutoff condition, which for $B > 0$ is given by $B < k_0^2$; this arises from the requirement that, in the NDR expression, the term $- \frac{1}{2}Z_r'(\frac{v_0}{\sqrt 2})$ must be negative for the fast Langmuir branch.
For $B < 0$, additional cutoff conditions arise—specifically $k_0^2 > \frac{-B}{2}$—stemming from the solvability condition for  Poisson's equation (i.e., no solution exists "below" the inverted electron hole (SPD)).

We conclude this section with a comparison. If we replace $B$ with $\lambda$ in (8) and (9), we arrive at van Kampen's linear analysis of undamped harmonic waves in the  linear Vlasov-Poisson (lVP) system, where $\lambda$ represents the continuous spectrum \cite{vanKampen55}. Thus, it is evident that our analysis constitutes the appropriate extension to the full nonlinear Vlasov-Poisson (VP) space, which also encompasses the structural profile extensions. 

\section{The periodic Langmuir hole}

 In the high-speed regime of Langmuir waves, we again consider the "off-dispersion" mode and find a solution for a given $B > 0$, provided the condition $0 < \sqrt{B} < k_0$ is met. This solution represents a specific cnoidal wave structure that exhibits a depression (a "hole" or "dip") in the region of trapped particles. It is thus the analogue of solitary holes based on the acoustic modes. However, an electron acoustic mode is absent here (note that $v_0 \rightarrow \infty$ when $k_0 \rightarrow \sqrt{B}$), which is why no Schamel-type evolution equation exists for this mode structure \cite{SPF25}.\\
This Langmuir hole is spatially periodic and must be distinguished from the better-known Langmuir soliton, the existence of which relies on the trapping of waves in a density depression \cite{S79b}.
In this limiting case, where $k_0^2$ approaches $B$ and the phase velocity consequently diverges (see the next section), the pseudo-potential $\mathcal V(\phi)$ takes the simple form $-\mathcal V(\phi)=\frac{B\psi}{2}\phi[1-(\frac{\phi}{\psi})^{3/2}]$; it is thus indeed a periodic structure rather than a solitary one. In the case of an additional trapping scenario $\Gamma$, the extended NDR condition (7) applies, and for a Langmuir hole, the condition $k_0^2 > B-\Gamma > 0$ must be satisfied, which constrains $\Gamma$.\\

It appears that this off-dispersion mode structure,  the  Langmuir hole, was observed in a recently published numerical experiment \cite{Ouyang26}. However, it requires the use of a modified expression for $B$. This modification arises because the background distribution function deviates from the Maxwellian distribution prior to the excitation of the structure and exhibits a quasilinear plateau in the resonance region.
The "regularized" electron hole proposed by Korn and Schamel \cite{KS96a} offers a necessary improvement in this regard. It relies on the fact that the first and second derivatives of $f_{0e}(v)$ with respect to velocity vanish at the resonance velocity. 
The corresponding B, denoted as $B_{KS}$, is given by
\begin{eqnarray}
    B_{KS}:=\frac{ 32 (-\beta)}{15(3+\tilde{v}_D^4)}\sqrt{\frac{\psi}{\pi}} e^{-\tilde{v}_D^2/2}.
\end{eqnarray}
and presents an updated expression for $B$ that corresponds to Equation (3.38) in \cite{KS96a}.
Fig. 1b of this paper shows that, in the case of a "regularized" electron hole, the deficit of trapped electrons persists but is reduced.\\

We see that $B_{KS} > 0$ requires the condition $\beta < 0$, which corresponds to a deepening (depression) of $f_e$ in the region of trapped particles, as observed in \cite{Ouyang26}.\\
We find that stronger regularization—specifically, one in which the second $v$-derivative of the total distribution $f_e$ also vanishes at the separatrix (given by $f_e''(v)=0$ at $\epsilon_e=0$)—would require $\beta=0$. This implies a complete lack of the trapping structure. 
Alternatively, we recover the undamped linear Langmuir wave that lies on the on-dispersion curve, noting that $f_{0e}'(v_0)=0$ holds.

But this is not what is observed in \cite{Ouyang26}. They have still a structure with a depressed trapped particle region.\\
The explanation for this apparent contradiction lies in the fact that higher-order regularization should be performed only in the immediate vicinity of the separatrix (specifically in a region—say, with a width of $|\epsilon_e| << \sqrt{|\epsilon_e|}$), where particle trajectories are irregular anyway \cite{TEXT1}.\\

In numerical methods, this is achieved through unavoidable numerical diffusion, whereas in real physics, it occurs through an opening of the VP framework in this region and the inclusion of higher-order correlations within the BBGKY hierarchy (or, for example, a Landau-type Fokker-Planck collision operator). \\
 In this context, we merely point out that the related question of anomalous resistivity was addressed in \cite{KS96b} (for mobile fluid ions) and in \cite{LS05} (for kinetic ions).

\section{Self-Acceleration due to Detrapping}

Another application of the off-dispersion mode $(k_0, B>0)$ and the Langmuir hole is the self-acceleration of a periodic electron hole—observed numerically by Lobo and Sayal \cite{Lobo25}—which is accelerated solely through interaction with electrons. Their model assumes stationary (immobile) background ions; consequently, the interaction with  ions proposed in the literature cannot be operative here.\\

To understand their result, we begin with the  periodic, slow electron acoustic hole that forms first and is designated by the index 1.

Assuming  $0<B<k_0^2<<1$ the NDR is solved by $v_{01}=1.307 (1 + k_0^2 - B)$.
After the acceleration, which itself defies description (including NDR) due to its transient behavior, the hole settles at the Langmuir solution (index 2) $v_{02}=\sqrt{\frac{1}{k_0^2-B} +3}$, where we used $-\frac{1}{2}Z_r'(\frac{v_0}{\sqrt 2}) \approx -\frac{1}{v_0^2} (1+ \frac{3}{v_0^2})$ valid  for $v_0>>1$. 

To see how the structural phase space is affected during acceleration we first note that there are hints that the structure itself mainly keeps its form, strength and scale \cite{SMS17, MSS18,MSS20, SMS20}. This implies that $B_1\approx B_2$ with $B$ given by
 (5)—with $\tilde v_D$ replaced by $v_0$. 
 It then holds\\
 $(1-\beta_2- v_{02}^2)\approx(1-\beta_1- v_{01}^2) e^{(v_{02}^2-v_{01}^2)/2}>>(1-\beta_1- v_{01}^2 )$
  or
 \\
 $-\beta_2 >> -\beta_1 +(v_{02}^2-v_{01}^2 >> -\beta_1 > 0$.

Consequently, $\beta_2$ is significantly more negative than $\beta_1$; this implies not only that a substantial reduction of the trapped population as a whole occurs during acceleration indicated by the exponential factor in (3), but also that the central region of the trapped electron distribution, in particular, is strongly depleted.

The origin of this acceleration (or this “detrapping” process) remains to be clarified. One argument posits that a faster hole possesses lower—typically negative—energy \cite{KS96b, LS05, S23}, thereby making free energy available for processes such as the emission of linear waves; this naturally accelerates the approach to thermal equilibrium in a collisionless plasma—that is, without the involvement of collisions.

We view this as an indication that the system possesses an inherent, potential mechanism (or tendency) to evolve toward negative-energy structures, and we attribute an attractor-like property to the Langmuir hole.\\
 \section{Failure of van Kampen superpositions}
 Before discussing the consequences further, we wish to directly demonstrate the absence of a relationship between linear VP dynamics and full VP dynamics. 
 When attempting to describe a cnoidal hole solution—specifically (77) from \cite{S00}—as a particular superposition of linear van Kampen modes, one encounters an infinitely long series that decays extremely slowly, even in the weakest nonlinear regime.

 In this solution, the convergence of the infinite series is achieved only through the alternating signs of the Taylor expansion terms; this is attributable to the contribution: $x\sin x$  in $\phi(x)$; see (81), (82) in \cite{S00}. Hence: "Only by a very delicate superposition of infinitely many modes a nonlinear solution could be constructed."

 In other words a single van Kampen mode and all other superpositions, except the just mentioned sophisticated one,  cannot be considered as a proper solution of the full nonlinear problem even in the infinitesimal amplitude limit $\psi \rightarrow 0^+$.

 Any arbitrary ordinary superposition is subject to phase mixing and Landau damping, without a persistent final structure remaining. The regimes of linear and nonlinear VP are decoupled and separated by a gap that cannot be bridged analytically.

 \section{The Cauchy Initial Value Problem, the Asymptotic Hole Selection Problem and the "Waterfall" Analogy}

 This brings us to the core topic of the study—a topic that practically imposes itself and leads to the "waterfall" analogy.\\ 
 The question is whether the Cauchy initial value problem for the VP system is uniquely solvable and—relatedly—whether a corresponding set of initial data exists for every Schamel solution. The examples, we have in mind,  are the linear two-stream instability and the nonlinear Landau damping; however, this also applies to other cases, such as the subcritical excitation of “holes” by initial “seeds,” as investigated numerically by Mandal, Sharma, and Schamel \cite{SMS17}, among others.
 
In my view, there are two approaches to this: one that operates strictly within the VP system—which I would characterize primarily as mathematical—and another that briefly steps outside the VP system, which I would describe as more physical and closer to reality, even though the methodology involved will be no less mathematically challenging (see later).\\

It is well known that the VP system is Hamiltonian—meaning that, under regular conditions, the flow in phase space can be determined by integration along the characteristics (particle trajectories). It is also known that the flow is constrained by so-called Casimir invariants, which, in a sense, keep it on a short leash. However, it is less clear whether, through the formation of separatrices, they allow for multiple incompatible distributions of trapped particles—and thus for the final Schamel states.
With the formation of a separatrix, a series of "irregular, anomalous" events emerges, which can be summarized under the following headings:

* logarithmically diverging residence (or transit) time of a particle on the separatrix

* folding of trajectories near the separatrix that spend time both outside and inside the separatrix

* associated with this is a sensitive dependence of the trajectories, where even the slightest initial deviations can trigger globally (macroscopically) perceptible changes in the flow

* accompanied by a filamentation of the flow that leads to increasingly smaller scales over asymptotic time and affects  fine-tuning, phase-mixing, coarse-graining etc. and resolvability of a structure.

* an accumulation (depletion) of trajectories as they migrate toward the separatrix, starting from the low- (high-) velocity side, which leads to a slope singularity of the form $|\partial f_e(x,v)/\partial v | \rightarrow \infty$ as $\epsilon_e \rightarrow 0^+$ (see Schamel solution with $v_0 \ne 0$)

* in case of the trapping scenario $\gamma$ in (3) one encounters a similar singularity $\sqrt{-\epsilon_e}$,
 in case of $\chi_{1,2}$ (see (1) in \cite{S23}) a singularity $\ln(-\epsilon_e)\sqrt{-\epsilon_e}$ and  $\ln(-\epsilon_e)^2\sqrt{-\epsilon_e}$, respectively, when approaching from the trapping side ($\epsilon_e \rightarrow 0^-$). These two latter trapping scenarios are responsible for the Gaussian profiles $\phi(x) \sim e^{-x^2}$ and $\phi(x) \sim e^{-\sinh(x)^2}$, respectively.

It is the singularity of the trajectory flow at the separatrix—where neighboring paths diverge—that leads to different trapping scenarios, even within deterministic Hamiltonian dynamics.
The flow loses its regular structure, in a sense; while it remains well-defined and measure-preserving as a Hamiltonian flow the mapping from initial data to the resulting trapping topology becomes singular.


We conclude this part with a comparison that practically forces itself upon us: the trapping process itself exhibits many characteristics of a \textbf{waterfall}.
Both defy mathematical description and thus represent a gap in the overall course of evolution. It is impossible to link the structures below the cascade—such as eddies or whirlpools—to the disturbances above it.

The situation is similar for collisionless plasmas: structures in the nonlinear Vlasov-Poisson (VP) regime—that is, those emerging after the trapping process—cannot be linked to processes occurring in the linear Vlasov-Poisson (lVP) regime prior to trapping.

The Hamiltonian flow remains smooth for any finite time $t$, yet the asymptotic classification of trajectories becomes singular at the separatrix; this results in non-integrability and the loss of an explicit formulation of the dynamics, accompanied by associated problems regarding the selectability  of equilibrium states.

An overview of the literature on Vlasov-Poisson dynamics in the context of structure formation—from the initial state to the final equilibrium—reveals that, to date, no single study has succeeded in demonstrating the existence of a Schamel equilibrium. The reason for this is simple: no study has been able to bridge the analytical gap.

The proof of nonlinear Landau damping provided by Mouhot and Villani \cite{MouhotVillani11} does not constitute a contradiction.


They have not solved the full fundamental problem of nonlinear Landau damping—which is linked to structure formation in the final stage of evolution.
They decided instead to identify situations in which the trapping process could be disregarded—that is, to position themselves above the cascade, to stick with the waterfall metaphor.
In fact, even slight (e.g. nonregular) deviations from their initial conditions are sufficient to trigger the trapping process and structure formation, as numerous numerical simulations show \cite{Manfredi97, LS05, SMS17,MSS18,MSS20,SMS20, Ouyang26}.\\

Hence, although electron holes have repeatedly been observed to emerge dynamically e.g. from unstable initial distributions in kinetic simulations, a first-principles derivation of a specific nonlinear hole equilibrium from a prescribed smooth initial state—i.e. a solution of the Vlasov–Poisson Cauchy problem that analytically determines the resulting trapping scenario and its Schamel parameters—appears to be lacking.

We therefore focus on two series of investigations that appear to support our undertaking:
those by Doveil, Elskens, Firpo, et al. \cite{Doveil2001, Firpo2001}, as well as those by Driscoll, Kabantsev, et al. \cite{Kabantsev2010, Kabantsev2014}.

Both are closer to the actual physical processes  by identifying the sensitivity of the dynamics near the separatrix and reveal the limitations of the Vlasov-Poisson description, highlighting the need for special measures to handle it. 

\section{Finite-(N) Dynamics, Separatrix Crossing
and chaotic separatrix crossing}
Doveil et al. \cite{Doveil2001, Firpo2001} investigated the dynamics of a finite number of particles in the presence of nonlinear electrostatic waves, with particular emphasis on particle trapping and separatrix dynamics. 
They found that finite-(N) effects produce a long-time behavior that differs from the numerical errors associated with Vlasovian approaches and relate this behavior to the pulsating crossing of the separatrix by individual particles.
Their result is significant because it identifies the separatrix not merely as a geometrical boundary in phase space, but as a dynamically active region in which the microscopic particle dynamics can have consequences for the subsequent macroscopic evolution.

Their results demonstrate that the transition between passing and trapped motion is a genuinely dynamical process rather than a simple adiabatic relabelling of particle orbits. Near the separatrix, the particle dynamics becomes strongly sensitive to perturbations and to the discreteness of the particle system. This provides a microscopic perspective on the transient dynamics underlying the formation of coherent phase-space structures.
 
In a later paper \cite{Escande2018}, the remarkable conclusion is drawn that the Vlasov limit is singular and corresponds to a renormalized description of the actual $N$-particle dynamics.
The question can therefore be made more precise: Is the separatrix the dynamical locus where the singularity of the Vlasov description manifests itself?
From this perspective, the separatrix would be a dynamical bottleneck—a concept that fits well with the "waterfall" metaphor used here.

However, these investigations do not formulate this process as a mechanism for selecting among the different trapped-particle distributions represented by nonlinear Vlasov-Poisson solutions nor do they formulate a BBGKY-based diffusion mechanism at the separatrix.

This latter step, which encompasses the effective dissipation brought about by chaotic separatrix crossings, was first proposed in \cite{Kabantsev2010, Kabantsev2014}.

Kabantsev et al. 
 demonstrated the importance of chaotic dynamics during separatrix crossing. Near the separatrix, small perturbations can produce substantial changes in individual particle trajectories, making the trapping process intrinsically sensitive and non-adiabatic. Such dynamics provides a natural link between finite-(N) particle motion and the mean-field Vlasov description: while the latter describes the smooth self-consistent state away from the microscopic crossing process, the detailed selection of trapped and passing populations occurs during a transient dynamical layer near the separatrix.

The separatrix, to summarize, may constitute a dynamically singular region in which the finite-N dynamics temporarily requires information beyond the closed Vlasov description, with higher-order correlations in the BBGKY hierarchy providing the natural microscopic framework for the associated phase-space mixing and subsequent coarse-graining.

These results are relevant to the present interpretation of trapping as a transient selection process. The Vlasov–Poisson equation describes the subsequent self-consistent evolution, but the microscopic mechanism determining which of the possible trapping scenarios is realized need not be contained in the idealized mean-field description itself. In this sense, separatrix crossing provides a natural interface between finite-(N) dynamics and the family of nonlinear trapped-particle states represented by Schamel-type solutions.

In this work, we have considered only the simplest case: stationary immobile ions and zero drift velocity. Accounting for these effects significantly broadens the spectrum of "holes" (phase-space holes). Together with other particle trapping scenarios \cite{S20,S23}, they lead to a much greater variety of such structures. This is of great importance, as this set includes negative-energy holes, which act as attractors in the evolutionary dynamics. In such a state, the total system possesses free energy; it is therefore more easily destabilized and reaches a higher level of turbulence more rapidly.
For further details, we refer to Luque and Schamel \cite{LS05} as well as Das, Borah, and Schamel \cite{DBS18} (and the references cited therein).\\
For a generalized application of this phase-space vortex analysis to inhomogeneous plasmas (e.g., drift-wave turbulence) or plasmas with a sheared magnetic field (e.g. fast magnetic reconnection), we refer to the review article by Eliasson and Shukla \cite{ES06}. 

\section{Hole selection via an asymptotic BBGKY separatrix boundary layer?}

We now return to the second—and, in our view, more physically grounded—possibility for dealing with the singularity problem.

This shifts the problem from classical Vlasov theory to the question of singularity removal through a higher level of kinetic description.

The argument is as follows: The Vlasov equation is precisely the collisionless limiting equation of the BBGKY hierarchy—that is, it arises from the assumption that the two-particle correlation $g_2$ vanishes or can be neglected in the corresponding limiting case. As long as the single-particle flow remains regular, this is consistent. However, the separatrix is precisely the region where this assumption fails. In other words, the Vlasov description is not globally incorrect; rather, it merely loses its validity within a thin boundary layer around the separatrix.

This is strongly reminiscent of the role of boundary layers in other asymptotic problems: the outer solution remains valid, yet a refined theory must be applied within a small region. In this problem of "matched asymptotic boundary values," two-particle correlation is significant only within an inner layer surrounding the separatrix—with a width of, say, $O(\epsilon_e) \ll O(\sqrt{\epsilon_e})$—whereas the outer region retains its Vlasov-Poisson character.

The selection operator would then be located exclusively within this inner layer.
This would offer a remarkable advantage: one would not need to solve the BBGKY hierarchy for the entire temporal evolution; it would suffice to understand its effect within a localized asymptotic layer.

In numerical methods, this occurs automatically—and to a certain extent uncontrollably—due to numerical diffusion. Effective diffusion is always present in PIC or Euler-Vlasov codes, whether caused by grid interpolation, finite particle numbers, or discretization. This diffusion smooths the separatrix and prevents the formation of infinitely steep gradients.

Numerical methods thus implicitly reproduce the physical effect of the neglected two-particle correlation—a remarkable fact. They may well yield the "correct" Schamel solutions, as their unavoidable diffusion compensates for the missing BBGKY hierarchy.

Naturally, caution is warranted: numerical diffusion is generally identical neither quantitatively nor necessarily qualitatively to the effective diffusion induced by correlations. Nevertheless, the analogy might point to a deeper connection.

Hole selection would then not be an inherent result of the Vlasov equation, but rather the outcome of a brief excursion to the next higher level of the hierarchy.

The mathematical research objective could thus be formulated as follows: Can it be shown that the separatrix forms an inner boundary layer in which the two-particle correlation—derived from the BBGKY hierarchy—becomes dominant, thereby triggering the selection mechanism of the Schamel trapping scenarios?

This would provide an elegant conceptual framework that does not refute Vlasov theory but rather refines its range of validity. The Vlasov equation describes the temporal evolution almost everywhere in phase space, whereas the actual selection of stationary trapping structures occurs within an asymptotically thin but dynamically crucial BBGKY boundary layer.

\section{Some further aspects and remarks}

First of all, it is evident that only Schamel’s pseudo-potential method meets the requirements for better identifying observed equilibrium structures, understanding numerical structural phenomena, or—going a step further—mathematically describing intermittent particle and energy transport.\\
In many respects, expansions and additions are possible (and necessary) to move closer to the goal. This involves the following aspects:\\

1. Develop greater confidence in handling undisclosed potentials that inevitably arise when the range of "trappings scenarios" is expanded.\\
2. A generalized derivation of the energy of a structured plasma for the purpose of optimizing energy reduction and the associated destabilization of the plasma.\\
3. Mathematical derivation of the stability of a structure as a function of various parameters, in order to determine whether they act as attractors within the dynamics.\\
4. The dynamics might also necessitate a return to the case of finite amplitudes, $\psi \simeq O(1)$, as employed in the initial phase of Schamel's theory \cite{S72,S79b,BS81,S82,SB83,S86}.\\
5. To extend the variety of solitary holes on double layers, only one additional condition is required: $\mathcal V'(\psi)=0$, provided that $\mathcal V'(0)=0$ already holds.\\
6. We point out that the problem addressed in this work bears a certain resemblance to the problem of plasma expansion into vacuum. A one-dimensional plasma initially confined to a half-space develops an accelerated, singular ion  front peak  over finite time; 
this frontal singularity arises from wave breaking, can be described by the Sack-Schamel equation, and can be stabilized by introducing diffusion in spatial space \cite{SS87}.\\
7. We wish to dispel a myth that has crept into the literature over time: the claim that Schamel's theory of ion holes places a rather stringent condition on the temperature ratio $\theta := \frac{T_e}{T_i}$. In fact, however, Schamel's theory does not entail such a restriction, as ion holes exist for all $\theta$ in the range $0 < \theta < \infty$, as was recently demonstrated in \cite{SPF25}.\\
8. It is to be hoped that this work initiates a process prompting the relevant segment of the plasma community to move away from the tendency to explain everything within the framework of Landau/BGK—that is, to force everything into the Landau/BGK straitjacket. Although the latter has been firmly established for half a century and has admittedly achieved some successes, it ultimately does not represent a viable alternative to our own approach.
\cite{S23,SC23,SPF25}.
Would you like an example?


The assertion made in \cite{VOND06}—which is presented quite casually, yet rests on a profound misunderstanding—that “due to the electrons trapped (in a BGK mode) the distribution is effectively flat at the phase velocity, and this turns off Landau damping,” proves upon closer inspection to be an overly broad and undifferentiated simplification; furthermore, it leads to conclusions that are simply incorrect.



The correct view is that Landau damping cannot play any role in the linear stability of hole equilibria, as it applies only to homogeneous equilibria and is therefore inapplicable. Moreover, such hole equilibria can be described only very incompletely by BGK theory, and—with the exception of single harmonic wave equilibria—no general analysis exists to date that can make a statement regarding general stability of hole equilibria. Moreover, a typical distribution of trapped particles is trough-shaped rather than plateau-like.
For holes, the Landau criterion is irrelevant, regardless of whether the situation is in the subcritical or supercritical regime. They are unaffected by it.
Furthermore, their term "electron acoustic wave", used later, should be updated and referred to as "slow electron acoustic wave" to avoid confusion and align with established hole notation.\\
9. We would like to point out that this explicit mathematical non-integrability during the “trapping gap” is further exacerbated by the mathematically open form of the potentials (undisclosed potentials)—resulting from the contribution of additional trapping scenarios \cite{S20,S23}.\\
10. Finally, it should be noted that maintaining ion immobility—given the inevitability of diffusion in phase space—would lead to the disappearance of all electron-hole structures in phase space \cite{KS96a,KS96b,LS05}. Only by lifting these constraints—that is, by accounting for realistic ion masses—is it possible to formulate suitable transport models for intermittent plasma turbulence in the regime of very long mean free paths \cite{SK96,LS05,HS94}; this regime is of interest for the operating range of future fusion facilities. However—it must be said—they presuppose a linearly unstable, two-stream like initial state.\\

 \section{An illustrative example}

Finally, a simple example serves to illustrate the overall situation and captures all essential elements, namely:
i) the numerical illustration of “trapping” as a decisive turning point in the evolution—followed by a phase of intense activity and, ultimately, a period of stabilization and settling, as well as the formation of a hole equilibrium—and
ii) the limited validity of the coupled-mode equations, which becomes apparent upon the onset of “trapping” and points to a gap in the analytical description.\\

It is the symmetric two-stream instability driven initially by two counter-propagating cold electron beams, as treated by Bauer and Schamel in \cite{BS92}. Figures 1 and 2 of that paper show that pattern or hole formation sets in precisely when the nonlinear terms in the coupled-mode equations begin to dominate the linear terms. 
This process unfolds quite spontaneously once the linearly most unstable mode reaches a value sufficient to trap the first electrons. This is followed by a cascade of modes that were previously damped but now also grow, leading to particle trajectories exhibiting highly irregular behavior.
Beyond this point, structure formation can no longer be traced analytically; moreover, it takes on a somewhat artificial character, as numerical diffusion acts as a smoothing mechanism and counteracts the physical excursion into the BBGKY hierarchy.

Consequently, one is confronted with the simultaneous excitation of an unlimited number of modes—a phenomenon that casts the concept of the “cascade” in a new light, as it can thus be applied to the trapping process itself rather than being understood merely metaphorically.

Bauer and Schamel thus demonstrate the existence of the dynamic transition but do not yet determine the mathematical selection mechanism by which a specific Schamel solution emerges from the multitude of possible trapped-particle scenarios.\\

\section{Additional Remark}
 And last but not least, we would not be surprised if electrostatic double layers played a role in the micro-shock experiments of \cite{Jepsen26}—which involve a capillary-confined plasma generated by an intense laser pulse—thereby tapping into a virtually limitless state space of double-layer-like hole structures.

\section{Summary and Conclusions}

A central difficulty in connecting the Vlasov–Poisson initial-value problem with the large variety of nonlinear coherent structures observed in collisionless plasmas is the selection of the trapping scenario. In a purely mean-field description, the Vlasov equation provides a deterministic evolution of the one-particle distribution function. However, the transition from an initially smooth (or less smooth) phase-space distribution to a state containing trapped particles involves the crossing of a separatrix, where the character of the single-particle dynamics changes qualitatively.

This suggests that trapping should not necessarily be regarded as an instantaneous event occurring entirely within the regular Vlasov–Poisson dynamics. Rather, it may be interpreted as a transient multiscale process in which the mean-field description becomes particularly delicate. In the vicinity of a separatrix, particle trajectories are extremely sensitive to small perturbations, and the usual adiabatic description breaks down. Consequently, the microscopic N-particle dynamics and the correlations represented by the BBGKY hierarchy may become relevant precisely during this short transition interval.

The important point is that the Vlasov equation is itself the mean-field limit of the BBGKY hierarchy. It is therefore not an isolated dynamical theory, but the leading-order description obtained when correlations between particles are neglected in an appropriate large-N limit. The trapping process raises the question of whether this reduction remains uniformly adequate during separatrix crossing. Even if the correlations remain small in an appropriate global norm, their local dynamical effect near the separatrix may become significant because the single-particle flow is singular in the sense of extreme sensitivity to initial conditions and of the divergence of characteristic time scales.

One may therefore envisage the following schematic sequence:

$\text{initial Vlasov state}
\longrightarrow
\text{separatrix crossing}
\longrightarrow
\text{transient correlated dynamics}
\longrightarrow
\text{selected trapping scenario}
\longrightarrow
\text{new effective Vlasov state}.$


The middle part of this sequence, with its cascading appearance, represents a gap in the analytical description that can be metaphorically viewed as a waterfall.

The system enters a short-lived regime in which different microscopic trajectories can experience different trapping histories and be the cause of a mathematical non-treatability.

After this transient stage, or black box in the description of the evolution, the system may again be adequately represented by an effective one-particle Vlasov–Poisson description, but now with a phase-space structure that contains a particular trapped-particle population.

This viewpoint offers a possible interpretation of the remarkable diversity of nonlinear Vlasov–Poisson structures. The multiplicity of Schamel trapping scenarios need not imply that the initial  distribution uniquely determines, through a simple regular evolution, one analytically predetermined trapped-particle distribution. Instead, the selection of the trapped population may be associated with the singular and multiscale dynamics of separatrix crossing. The resulting state can subsequently evolve on the Vlasov–Poisson level and may be represented by one of the corresponding nonlinear Schamel solutions.

In this interpretation, the role of the BBGKY hierarchy is not to replace the Vlasov–Poisson description, but to provide a possible dynamical framework for the otherwise unresolved transition between an untrapped and a trapped phase-space configuration. The higher-order correlations would then be relevant primarily during the transient crossing process, while their detailed microscopic information becomes progressively coarse-grained in the subsequent mean-field description.

This also clarifies the conceptual role of Schamel-type solutions. They may be regarded not merely as stationary mathematical solutions constructed by prescribing a trapped-particle distribution, but as possible asymptotic representatives of states reached after a complicated trapping transition. The nonlinear dispersion relation then expresses the self-consistency of the resulting structure, while the particular trapping scenario specifies the phase-space population produced by the preceding transient dynamics.

A rigorous realization of this picture would require a non-uniform analysis of the mean-field limit in the vicinity of separatrix crossing. In particular, one would have to determine whether correlations neglected in the ordinary Vlasov limit acquire a dynamically enhanced influence in the separatrix region and how this influence is subsequently transferred into the coarse-grained one-particle distribution. Such an analysis could provide a mathematical bridge between N-particle dynamics, the BBGKY hierarchy, separatrix crossing, and the multitude of nonlinear Vlasov–Poisson trapping states.

The essential conjecture is therefore not that the Vlasov equation fails, but that \textbf{the passage through the trapping transition may contain information that is effectively lost when the N-particle dynamics is reduced immediately to a smooth mean-field description}. The diversity of nonlinear coherent structures could then be understood as the macroscopic trace of different microscopic trapping histories.

In summary, the emergence of a separatrix in the wave structure of a Vlasov–Poisson (VP) plasma poses a fundamental problem for the description of nonlinear structure formation. Although the Hamiltonian particle flow remains well defined at finite times under appropriate regularity conditions, the asymptotic classification of trajectories becomes singular near the separatrix, complicating the description of particle trapping and the selection of the resulting coherent structures.

We propose that the role of particle correlations, as represented by the BBGKY hierarchy beyond the VP mean-field description, deserves particular attention during this transient phase. Correlation-induced smoothing and subsequent coarse-graining may provide a mechanism for matching the dynamics near the separatrix to a post-transient VP description. In this framework, matched asymptotic expansions offer a possible route to investigating how the local dynamics near the separatrix affect the selection of trapping scenarios and the associated Schamel equilibria.

A separatrix, therefore, does more than just distinguish between free and trapped particles; it also serves as a dividing line between linear Vlasov-Poisson dynamics and the actual Vlasov-Poisson dynamics that emerge following the transient phase. 
The latter is characterized by a temporary excursion into the correlation dynamics, which produces a smoothing effect in the vicinity of the separatrix.
 Stationary structures are situated within this second VP-regime.
 From this perspective, the trapping of particles—conceived as a kind of “waterfall” within Vlasov-Poisson dynamics—takes on a new significance.
 
Undoubtedly, resorting to the BBGKY hierarchy—in order to account for correlation effects—is also necessary for smoothing out singularities elsewhere in phase space, such as those arising from filamentation ("coarse-grained" averaging).

Whether this approach can ultimately help to clarify the role of trapping scenarios—and, in particular, the selection of specific scenarios—remains to be seen.
In a real plasma, fluctuations exert an influence not only at the beginning but throughout the entire course of its development.

\section*{Acknowledgements}
Portions of the manuscript were developed with the assistance from ChatGPT (OpenAI) and subsequently verified and edited by the author.
\section{Author Declarations}
The author has no conflict to disclose.
\section{Data Availability Statement}
Data sharing is not applicable to this article as no new data were created or analyzed in this study.


\begin{thebibliography}{99}
\bibitem{TEXT1}
This could shed light on theoretical studies that predict a threshold amplitude for the existence of a structure, as they may rely on distributions that are too smooth (regular).
\end{thebibliography}
\end{document}